# Generating Unseen Nonlinear Evolution in Sea Surface Temperature Using a Deep Learning-Based Latent Space Data Assimilation Framework

Qingyu Zheng[a], Guijun Han[a,*], Wei Li[a,*], Lige Cao[a], Gongfu Zhou[a], Haowen Wu[a], Qi Shao[b], Ru Wang[a], Xiaobo Wu[a], Xudong Cui[a], Hong Li[a], Xuan Wang[a]

[a] *Tianjin Key Laboratory for Marine Environmental Research and Service, School of Marine Science and Technology, Tianjin University, Tianjin 300072, China*

[b] *Fujian Key Laboratory on Conservation and Sustainable Utilization of Marine Biodiversity, Minjiang University, Fuzhou 350108, China*

\* Corresponding author.

E-mail address:

3019227010@tju.edu.cn (Q. Zheng), guijun_han@tju.edu.cn (G. Han), liwei1978@tju.edu.cn (W. Li), clg@tju.edu.cn (L. Cao), zhougongfudlmu@163.com (G. Zhou), whw@tju.edu.cn (H. Wu), shaoqi@tju.edu.cn (Q. Shao), wangruu@tju.edu.cn (R. Wang), xb_wu@tju.edu.cn (X. Wu), marine_cui@tju.edu.cn (X. Cui), hongli@tju.edu.cn (H. Li), xuanwang@tju.edu.cn (X. Wang).

**Highlights**

- We focus on capturing and generating unseen nonlinear evolution in the Earth system.
- We propose a novel data assimilation (DA) framework for multi-source data fusion.
- It provides a high-performance deep learning paradigm for DA in latent space.
- It can extract fine nonlinear structures from limited information.
- It can generate physically interpretable multiscale evolution.

**Abstract**

Advances in data assimilation (DA) methods have greatly improved the accuracy of Earth system predictions. To fuse multi-source data and reconstruct the nonlinear evolution missing from observations, geoscientists are developing future-oriented DA methods. In this paper, we redesign a purely data-driven latent space DA framework (DeepDA) that employs a generative artificial intelligence model to capture the nonlinear evolution in sea surface temperature. Under variational constraints, DeepDA embedded with nonlinear features can effectively fuse heterogeneous data. The results show that DeepDA remains highly stable in capturing and generating nonlinear evolutions even when a large amount of observational information is missing. It can be found that when only 10% of the observation information is available, the error increase of DeepDA does not exceed 40%. Furthermore, DeepDA has been shown to be robust in the fusion of real observations and ensemble simulations. In particular, this paper provides a mechanism analysis of the nonlinear evolution generated by DeepDA from the perspective of physical patterns, which reveals the inherent explainability of our DL model in capturing multi-scale ocean signals.

## 1. Introduction

As a key component of the Earth system, the ocean plays a vital role in regulating global climate and maintaining ecological balance. Accurate prediction of the multi-scale evolution of the ocean is our common goal [1]. However, the nonlinear characteristics of the ocean pose a major challenge to traditional data assimilation and prediction systems. In recent years, observation technology has developed rapidly and data has become increasingly abundant. This brings a new opportunity to deeply understand nonlinear evolution such as extreme weather and ocean processes. We also have higher requirements for the accuracy of Earth system predictions. The uncertainty in numerical predictions of the Earth system is largely controlled by the quality of initial conditions [2]. To solve this problem, data assimilation (DA) technology has been developed and applied in the optimization of initial conditions. DA is a method that obtains an optimal estimate [3] of the initial state by fusing multi-source information (observations and simulations). Among them, observations can represent the relatively real system state, but the distribution of observations is sparse [4]. The model simulations can represent the system state completely, but model errors make the simulation results often inaccurate [5]. Therefore, DA is generally considered to be very effective in improving the quality of initial conditions.

A series of studies have been devoted to the update and application of DA technology, summarized into two categories [6,7]: variational DA methods and filtering DA methods. Typical variational DA includes three-dimensional variational (3D-Var) and four-dimensional variational (4D-Var) DA methods [8,9]. Variational DA is often employed to solve complex optimization problems, so large amounts of computing resources are required. Most of the current filtering DA methods employ the Kalman filter or its variants [10–13]. Filtering DA is suitable for the fusion of real-time observations, but it requires model simulation to quantify the covariance structure before assimilating observations [14,15]. In summary, the above methods all require running the model repeatedly, so the traditional DA methods are computationally expensive and time-consuming [16]. In addition, sparse observations also prevent the fine structures in nonlinear evolution from being seen. Therefore, improving data fusion efficiency and paying attention to unseen nonlinear evolutions are major challenges facing current DA methods[17–19].

Recently, deep learning (DL) technology has been applied in many fields, such as computer vision [20], natural language processing [21] and image or text generation [22]. In Earth science, DL has also been successfully applied in the fields of oceanography, meteorology and remote sensing to help humans understand scientific issues of the Earth system [23–26]. A large number of studies have emphasized that the combination of DL and DA will bring breakthroughs in numerical prediction [27–30]. In fact, DL has been involved in some specific tasks of DA. For example, the DL model is combined with the Weather Research and Forecasting Model (WRF) to learn the DA process of 3D-Var [31], which is driven by observations to generate analysis increments. Furthermore, DL is effective in estimating model bias [32], which is close to weakly constrained 4D-Var. In order to take full advantage of the automatic differentiation of the DL framework, DL models are employed to construct tangent linear models and adjoint models in variational DA [33]. To speed up the DA process, a multi-layer perceptron (MLP) is employed to learn the relationship between observations and model solutions [34], which can learn model parameters by minimizing the mean square error between the MLP and 4DVar results. Similarly, a recurrent neural network [35] is trained to learn DA, which takes the distance between numerical predictions and analysis results as a constraint. In addition, an "End-to-End" DL framework is designed as an efficient solver for DA problems [36], which performs well with the help of automatic differentiation. Overall, different DL methods have been applied to various sub-problems of DA, including spatio-temporal interpolation [37], downscaling [38] and parameter estimation [39]. These results are encouraging, however, there are still three issues that need a further breakthrough.

First, the dimension of ocean state is usually large, which directly affects the efficiency of multi-source data fusion. DL models perform well in low-dimensional systems, but oceanic systems are actually much more complex. In fact, the power of DL models is feature extraction and compression. Therefore, DA in the original state space is not the optimal choice for DL models. We draw inspiration from a study [40] on latent space data assimilation using neural networks (NNDA), which has been applied to the assimilation of atmospheric states. The results show that it cannot only improve efficiency but also help DL better understand and represent nonlinear processes in dynamic systems.

Second, we focus on the nonlinear and multiscale evolution of the ocean, which is generally more difficult to observe than the atmosphere. The internal structure and dynamic changes of the ocean are more complex and difficult to capture directly through conventional observation methods. Latent space DA, such as NNDA, performs well in capturing atmospheric states, but the scheme design and performance evaluation of ocean processes are still insufficient. In addition, ocean observations of different resolutions have a significant impact on the stability of DL models. Therefore, generating unseen nonlinear evolution only through known observations and background information requires us to develop a "End-to-End" deep learning DA framework for heterogeneous data.

Third, deep learning models are considered a "black box". The lack of explainability has become a huge obstacle to the promotion of data-driven models in earth science research. We hope that the generated nonlinear evolution is reasonable, which requires the underlying physical connections between the data to be transparent. At the same time, our framework should be compatible with data from different sources (simulations, reanalyses or observations), which also requires a consistent physical constraint that can reveal the generation of nonlinear evolution.

To address the issues mentioned above, we propose a new explainable latent space data assimilation framework (referred to as DeepDA) based on deep learning for generating unseen nonlinear evolutions in the ocean. We draw inspiration from existing methods, including linear order reduction schemes [41,42] and sequential DA schemes under simplified models [43]. First, we design a generative proxy model (GenPM) based on self-supervised learning that can map the ocean state into a latent space. It is worth mentioning that we build a spatio-temporal attention residual (STAR) module and integrate the STAR module into the GenPM, which can help the DeepDA extract spatio-temporal multi-scale features. In order to improve scalability, the GenPM and latent space DA in the DeepDA are designed as independent modules. We employ the 3D-Var constraint form to construct the loss function of latent space DA. The latent space DA module employs automatic differentiation techniques in the DeepDA framework and is compatible with different tasks. Such a design can not only meet the needs of multi-modal data fusion, but also improve the overall DA efficiency. In order to verify the feasibility and explainability of DeepDA, we will conduct evaluation and analysis experiments using sea surface temperature (SST) as an example. To sum up, the major contributions of this study are summarized as follows:

- We devise a novel purely data-driven DA framework (DeepDA) in latent space, which can effectively fuse multi-source information and generate unseen nonlinear evolution. The DeepDA is guided by general variational constraints, which achieves superior performance in different tasks.

- We design an advanced GenPM to learn the nonlinear evolution of SST. The STAR module is designed to boost the ability of the GenPM to capture nonlinear features and outperform existing baseline models.

- We evaluate the robustness of DeepDA in capturing the multi-scale evolution of SST, and analyze the impact of factors such as observation resolution and ensemble background information. The performance of DeepDA has been well verified in real observations.

- We explore the physical explainability of DeepDA from a pattern perspective. The spatio-temporal evolution of the El Niño/Southern Oscillation (ENSO) can be explicitly captured from our deep learning model. The physical properties of the latent patterns can better guide DeepDA to generate refined nonlinear structures.

The rest of this article is organized as follows. Section 2 introduces the design details of the DeepDA framework and the experimental design of our work. Section 3 shows the experimental results and evaluates the performance of DeepDA. Section 4 discusses the robustness and physical explainability of DeepDA. Finally, Section 5 concludes this article.

## 2. Methods

### 2.1. Study Region and Variable

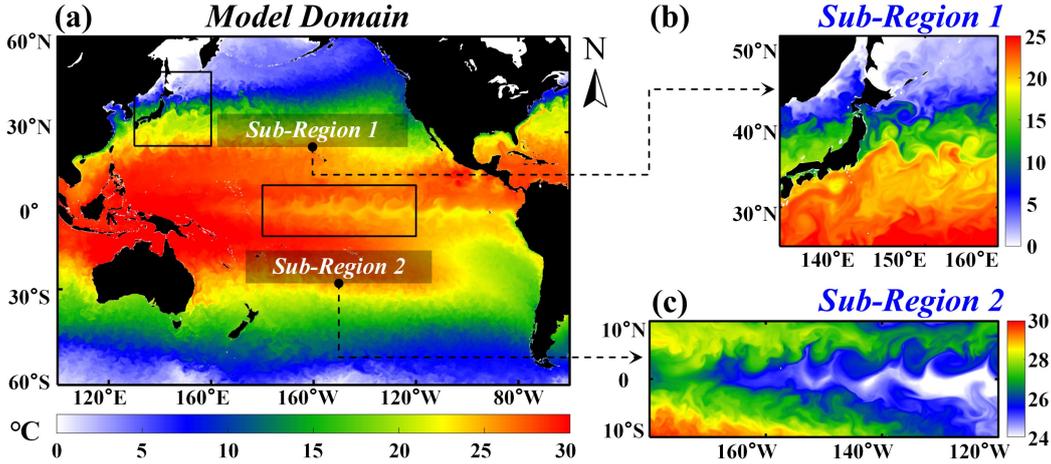

**Fig. 1.** Study region and experimental variable: (a) sea surface temperature (SST) in the model domain (Pacific Ocean), sub-regions of (b) the northwestern Pacific and (c) the equatorial Pacific. (Date: January 1, 2021; Source: Copernicus Marine Data Store).

As shown in Fig. 1(a), the study region (model domain) we focus on is the core area covering the Pacific Ocean (60°S~60°N, 100°E~60°W), which is affected by complex physics and extreme nonlinear processes (short-term meteorological effects, long-term climate change, and so on). Among them, the El Niño/Southern Oscillation (ENSO) is the strongest interannual climate fluctuation [44]. In particular, the Niño 3.4 region (the core region for studying the El Niño phenomenon) spans the east-central equatorial Pacific between 170°W~120°W, 5°N~5°S. In addition, the two sub-regions in Fig. 1 are intended to show more details of the nonlinear evolution.

It is well known that sea surface temperature (SST) is an important observed and predicted variable that affects atmosphere-ocean interactions. In addition, SST products (observation or reanalysis) are relatively mature, which is beneficial to training and testing our DL model. Therefore, in order to evaluate the performance of the DeepDA, we conduct observing system simulation experiments and multi-source data fusion experiments for SST.

### 2.2. Latent Space Data Assimilation Framework Based on Deep Learning

#### *2.2.1. Preliminaries*

The core idea of the DeepDA is to perform DA in a low-dimensional latent space. However, it is difficult to completely represent nonlinear evolution using linear methods. Therefore, how to design a DL model to capture the nonlinear latent space is the key to the DeepDA. The DeepDA employs self-supervised DL to build the generative proxy model (GenPM), which enables it to autonomously extract multi-scale features from large datasets and map them into a low-dimensional latent space.

$$\mathbf{z} = encoder(\mathbf{x}) \tag{1}$$

$$\hat{\mathbf{x}} = decoder(\mathbf{z}) \tag{2}$$

The proxy model consists of two components: encoder and decoder. Eq. (1) shows that the encoder of the proxy model maps the original state $\mathbf{x}$ to a low-dimensional latent vector $\mathbf{z}$. The decoder in Eq. (2) is designed as a mirror structure of Encoder, which can map the latent vector $\mathbf{z}$ back to the original state $\hat{\mathbf{x}}$. It should be noted that an ideal (error-free) proxy model will reconstruct the perfect state ($\hat{\mathbf{x}} = \mathbf{x}$). This model structure is often called Auto-Encoder (AE) in classic DL. AE can achieve low-dimensional mapping, but the features represented by the latent space are generally discrete. In practice, the distribution of states is non-discrete, which requires the proxy model to capture the continuous representation of nonlinear features in the latent space. Therefore, the DeepDA employs an additional random sampling mechanism in the encoder-decoder architecture, which can replace discrete latent features with probability distributions. The learning process of the proxy model can be simplified into an optimization problem in Bayesian theory:

$$\max_{\theta} p(\mathbf{x}) = \max_{\theta} \int_{\mathbf{z}} p_{\theta}(\mathbf{x}|\mathbf{z}) p(\mathbf{z}) d\mathbf{z} \tag{3}$$

where $p(\cdot)$ is the probability density function and the mapping relationship $p_{\theta}(\mathbf{x}|\mathbf{z})$ from latent vector $\mathbf{z}$ to state $\mathbf{x}$ is controlled by model parameter $\theta$. In self-supervised learning, the optimal parameter $\theta$ will be learned, which can satisfy the self-probability maximization matching of state $\mathbf{x}$. In other words, a known distribution of latent vectors and optimal parameters jointly determine the performance of state reconstruction. At the same time, the encoder can be expressed as:

$$q_{\phi}(\mathbf{z}|\mathbf{x}) \approx \frac{p_{\theta}(\mathbf{x}|\mathbf{z}) p(\mathbf{z})}{p_{\theta}(\mathbf{x})} \tag{4}$$

where $q_{\phi}(\mathbf{z}|\mathbf{x})$ is the conditional probability density function mapping relationship from latent vector $\mathbf{z}$ to state $\mathbf{x}$ controlled by model parameter $\phi$. Eq. (4) is the approximate posterior distribution of the latent vector, which shows that the mapping relationship between encoder and decoder is a continuous probability density distribution under parameter constraints.

In order to learn better parameters and distributions, it is necessary to design a reasonable training loss function. As shown in Eq. (5), the loss function consists of two parts. Among them, $Loss^{rec}$ can evaluate the reconstruction performance of state $\mathbf{x}$ (reconstruction loss) and $Loss^{reg}$ can constrain the distribution of the latent vector $\mathbf{z}$ (regularization loss).

$$Loss = Loss^{rec}(\mathbf{x}, \hat{\mathbf{x}}) + Loss^{reg}(\mathbf{x}, \mathbf{z}) \tag{5}$$

In the assumption of Gaussian distribution, $Loss^{rec}$ can be written as:

$$Loss^{rec}(\mathbf{x},\hat{\mathbf{x}}) = Loss_\theta^{rec}(\mathbf{x}) = -\log p_\theta(\mathbf{x}|\mathbf{z}) \tag{6}$$

In order to optimize parameters, the proxy model in the DeepDA employs the form of Huber Loss function to rewrite Eq. (6) as:

$$Loss^{rec}(\mathbf{x},\hat{\mathbf{x}}) = \omega \langle H_\delta(\mathbf{x},\hat{\mathbf{x}}) \rangle \tag{7}$$

$$H_\delta(\mathbf{x},\hat{\mathbf{x}}) = \begin{cases} \frac{1}{2}(\mathbf{x}-\hat{\mathbf{x}})^2, & if\ |\mathbf{x}-\hat{\mathbf{x}}| \leq \delta \\ \delta|\mathbf{x}-\hat{\mathbf{x}}| - \frac{1}{2}\delta^2, & if\ |\mathbf{x}-\hat{\mathbf{x}}| > \delta \end{cases} \tag{8}$$

where $\omega$ represents a weight coefficient greater than 0, which can control the balance of the two loss terms. $\langle \cdot \rangle$ represents the average of all grid points and $\delta$ is set to 1. To make $p(\mathbf{z})$ conform to the standard normal distribution, $Loss^{reg}$ is written as:

$$Loss^{reg}(\mathbf{x},\mathbf{z}) = -\log p(\mathbf{z}) + \log q_\phi(\mathbf{z}|\mathbf{x}) \tag{9}$$

During the training of the proxy model in the DeepDA, the encoder can generate matching vectors of mean ($\mu$) and log-variance ($\log \sigma^2$), representing the multivariate Gaussian distribution. Each latent vector element is then randomly sampled from this distribution [45]. In multi-source information fusion, the latent space changes slightly. So, the design of sampling and regularization contributes to the diversity of state mapping. At the same time, the continuity of adjacent latent spaces can ensure the similarity of decoding states. Therefore, mapping results are similar to ensemble perturbations of state variables under consistent physical constraints.

### *2.2.2. Design of Generative Proxy Model (GenPM)*

To enhance the scalability of the DeepDA framework, we develop a new GenPM named spatio-temporal attention variational autoencoder (STAVAE). As shown in Fig. 2(a), the STAVAE adopts a convolutional encoder-decoder structure. Overall, the STAVAE contains five modules: input, convolutional encoder, latent space mapping, convolutional decoder, and output. Among them, in order to extract deeper features, we independently design a spatio-temporal attention residual (STAR) module.

First, the encoder captures different scale features from state variables and forms downscaled feature maps. With the strengthening of the convolution channel and the reprocessing of the attention mechanism, the extracted nonlinear information is richer. Second, the latent space mapping module generates a distribution of states from the encoded feature map, which is compressed into a low-dimensional vector by a fully connected layer. Finally, the decoder restores the scale of the feature map until it reconstructs an output with the same dimensions as the input variables. In practice, the spatial information of variables is usually described on the latitude and longitude grid. Therefore, the design of the STAVAE model meets the needs of grid fusion and pixel-level scalability of multi-source data.

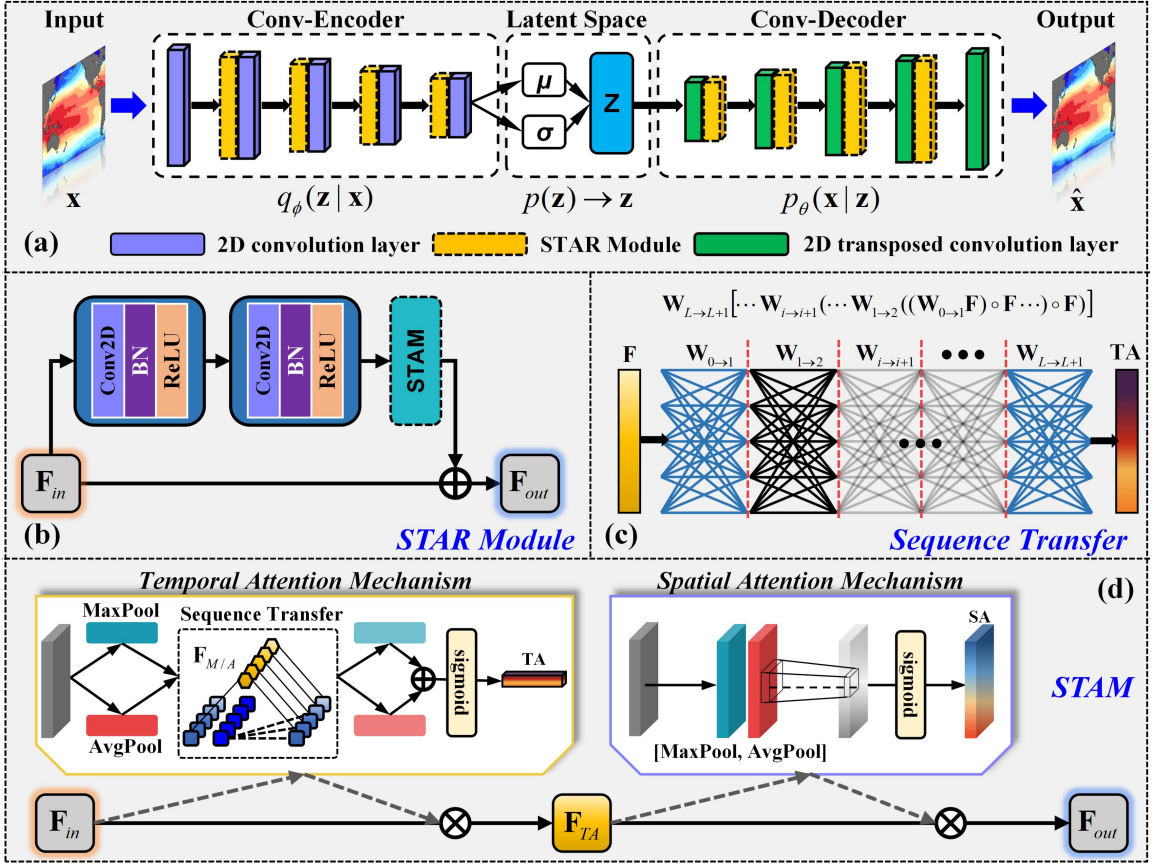

**Fig. 2.** Overall (a) and design details of our GenPM model (STAVAE): (b) the spatio-temporal attention residual (STAR) module, (c) the sequence transfer mechanism and (d) the spatio-temporal attention mechanism (STAM) in STAR module.

In the STAR module (Fig. 2(b)), the residual structure and spatio-temporal attention mechanism (STAM) are effectively integrated to achieve multi-scale feature extraction. The residual structure consists of two stacked convolutional neural network (CNN) layers and a STAM, where batch normalization (BN) and activation function (ReLU) are bundled with a 2D-CNN layer. In the training of the STAR module, the attention weights of different signals are learned to capture the dominant signals of evolution, which is beneficial to the feature generation of the latent space. STAM is an improved version of the Convolutional Block Attention Module (CBAM). CBAM [46] is a simple and effective attention module. It can be integrated into any CNN architecture with negligible overhead. It is worth mentioning that CBAM was originally designed for image segmentation or classification, so its channel attention module employs a simple multi-layer perceptron (MLP) to generate channel attention weights. However, in our task, the channel features represent continuous evolution, which needs to emphasize the sequential dependence relationship. Therefore, we design a sequence transfer mechanism [47] to improve the feature extraction performance of MLP in CBAM, forming a STAM. The sequence transfer mechanism weights the input channel feature map layer by layer, which can extract internal sequence dependencies while retaining all information of channel features. In fact, sequence transfer employs multiple unbiased fully connected layers with the same width as the feature dimension, which is close to the computational efficiency of MLP. Therefore, the STAR module can effectively improve the spatio-temporal representation capability of CNN and is computationally friendly.

In detail (Fig. 2(a)), the encoder and decoder in STAVAE each contain 5 convolutional layers. Each layer employs a 3×3 convolution kernel for feature extraction, in which the feature dimension will be reduced to half of its original size and the number of channels will be doubled. The mean and variance of the latent distribution are calculated from the encoded feature map and converted into a low-dimensional latent space vector. In addition, the STAR module only extracts spatio-temporal dependencies of features without changing the dimensions.

### *2.2.3. Framework details of DeepDA*

As mentioned before, the DeepDA contains two parts: the GenPM (Section 2.2.2) and latent space DA. In this section, we customize the framework details of DeepDA based on 3D-Var (Fig. 3(a)). 3D-Var seeks the optimal information fusion of background and observation by minimizing the cost function. The cost function has the following form:

$$J_{3D}(\mathbf{X}) = \frac{1}{2}\|\mathbf{X} - \mathbf{X}_b\|^2_{\mathbf{B}^{-1}} + \frac{1}{2}\|\mathbf{y} - H(\mathbf{X})\|^2_{\mathbf{R}^{-1}}, \tag{10}$$

where $\mathbf{X}$ is the analysis state, $\mathbf{X}_b$ and $\mathbf{y}$ are the background and observation respectively. $\mathbf{B}$ and $\mathbf{R}$ represent the error covariance matrices of background and observation. $H(\cdot)$ is the projection operator of observation, which can project the state into the observation space. $\|\cdot\|^2_{\mathbf{\Phi}^{-1}} = (\cdot)^T \mathbf{\Phi}^{-1}(\cdot)$ represents the Mahalanobis norm in which the $\mathbf{\Phi}$ means the covariance matrix. By minimizing the cost function, we obtain the optimal analysis state $\mathbf{X}_a$.

In traditional DA, the main challenge in minimizing the cost function is that the dimension ($\approx 10^{18}$) of the $\mathbf{B}$ matrix is too large, which makes the calculation of the cost function and gradient very difficult. In the DeepDA framework, the background state is defined in the latent space rather than the grid space, which still satisfies the Gaussian distribution assumption of background and observation errors in 3D-Var. Therefore, the latent space cost function of DeepDA is expressed as follows:

$$J_{3D-DeepDA}(\mathbf{z}) = \frac{1}{2}\|\mathbf{z} - \mathbf{z}_b\|^2_{\mathbf{B}_\mathbf{z}^{-1}} + \frac{1}{2}\|\mathbf{y} - H(D(\mathbf{z}))\|^2_{\mathbf{R}^{-1}}, \tag{11}$$

where $D(\cdot)$ represents the decoder of the GenPM, and $\mathbf{B}_\mathbf{z}^{-1}$ is the error covariance matrix of the latent vector with $\mathbf{D}$ representing the tangential operator of decoder $D(\cdot)$. The observation operator $H(\cdot)$ employs bilinear interpolation to project the decoded state into the observation space. Eq. (11) can directly measure the distance of the latent vector between the analysis $\mathbf{z}$ and the background $\mathbf{z}_b$. After the cost function is minimized, we can obtain the optimal latent analysis vector $\mathbf{z}_a$. It should be noted that the DeepDA transforms the 3D-Var minimization problem into a latent space, which greatly reduces the cost of gradient optimization. Taking our study as an example, a single state vector $\mathbf{X}$ has 480×800=384000 elements, and the original $\mathbf{B}$ matrix has more than $10^{11}$ elements. But in our DeepDA, the latent vector $\mathbf{z}$ has 128 elements, so the matrix $\mathbf{B}_z$ only contains about $10^4$ elements, which can be easily calculated by a personal computer. It can be found in the experiment that $\mathbf{B}_z$ is a diagonally dominant matrix, and the diagonal elements are several orders of magnitude larger than the off-diagonal elements (see Appendix A).

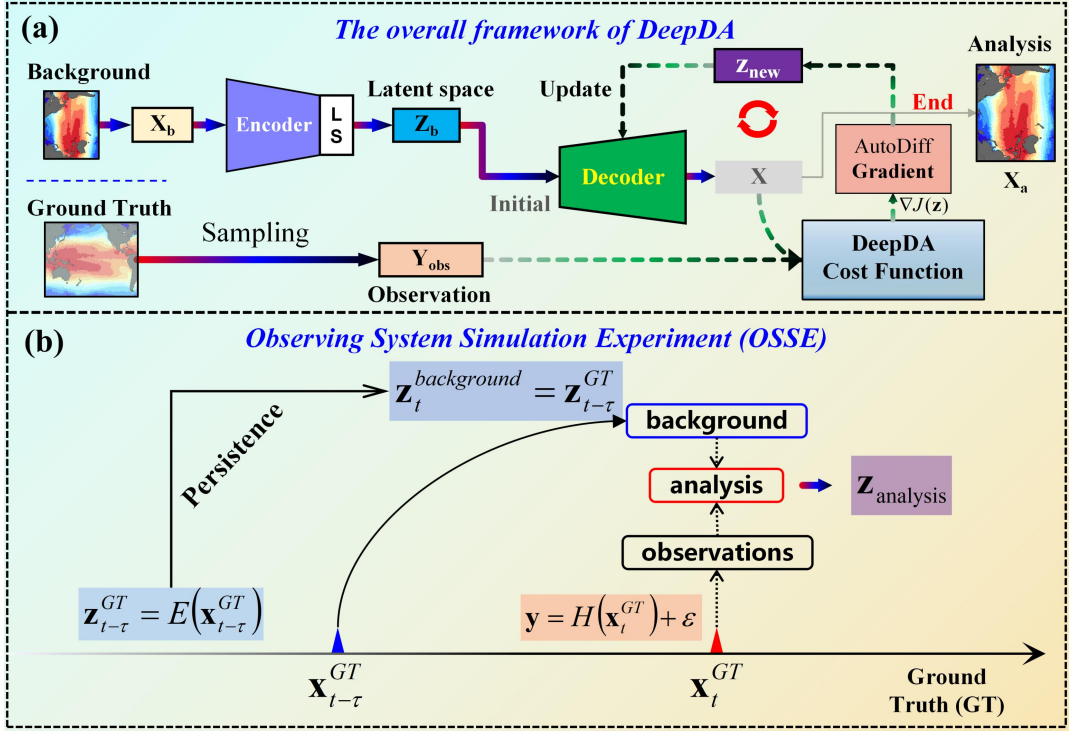

**Fig. 3.** End-to-end working pipeline of the DeepDA: (a) the overall framework of DeepDA and (b) the flowchart of the observing system simulation experiment (OSSE) with DeepDA.

2.3. Experimental Setup

*2.3.1. Data Preparation*

To train and evaluate the performance of the DeepDA framework, we employ daily average data of sea surface temperature (SST) as the state variable for the assimilation experiment. Hourly SST data are provided by the ERA5 (the fifth generation ECMWF atmospheric reanalysis product) [48], released by the European Centre for Medium Weather Forecasting (ECMWF), with a spatial resolution of 25 km and a regular latitude and longitude grid. We convert hourly SST data into daily averages. In order for STAVAE to produce a standardized input and output, we subtract the 30-year climate average of SST and normalize the SST anomalies. Our study region covers 480×800 grid points. The temporal coverage is 43 years, from 1981 to 2023. Among them, data from 1981 to 2010 (30 years in total) are used to calculate climate averages and data from 1989 to 2018 are selected to train the GenPM. In addition, the data from 2019 is used to evaluate the performance of different GenPMs and the data from 2020 to 2023 are used for the observing system simulation experiment (OSSE, Section 2.3.2) of DeepDA.

In order to explore the practical application potential of the DeepDA framework, we also design a series of multi-source data fusion experiments (Section 2.3.3). Different from OSSE, the objects of multi-source data fusion experiments use real observations and model simulations (background information). We employ Optimum Interpolation Sea Surface

Temperature (OISST) product from National Oceanic and Atmospheric Administration (NOAA) as observations, which integrate observations from different platforms (satellites, ships, buoys and Argo floats) into a regular global grid. In order to evaluate the robustness of DeepDA in different sample (ensemble) background information, we select the current best numerical model, the ECMWF's Seasonal Forecasting System (SEAS5), as the background. We extract the monthly SST ensemble prediction results of SEAS5 [49] from 2020 to 2023, which contains a total of 51 ensemble members. As before, we need to interpolate the SEAS5 (1°) onto the ERA5 grid (0.25°), which ensures the consistency of the data structure in the DeepDA.

### 2.3.2. Observing System Simulation Experiments

In order to capture the evolution details of SST, the DeepDA focuses on sea surface temperature anomalies (SSTA). We design a series of OSSEs to evaluate the effectiveness of the DeepDA. For simplicity, a commonly used baseline method (persistence prediction) is employed to generate the background fields. In Fig. 3(b), we show the details and flowchart of OSSE, where observations and background are derived from the ERA5 ground truth. We employ bilinear interpolation to project the ground truth onto the observation points and add random Gaussian perturbations to obtain pseudo-observations (Eq. (12)).

$$\mathbf{y} = H(\mathbf{x}_t^{GT}) + \varepsilon \tag{12}$$

In all OSSEs, the encoder $E(\cdot)$ in the GenPM has been employed to convert the persistent (lead time $\tau$ days) ground truth $\mathbf{x}_{t-\tau}^{GT}$ into background latent space vectors (Eq. (13)).

$$\mathbf{z}_t^{background} = \mathbf{z}_{t-\tau}^{GT} = E\left(\mathbf{x}_{t-\tau}^{GT}\right) \tag{13}$$

Based on the generated background and observations, the DeepDA will minimize the cost function and obtain the analysis result. In practice, the persistence lead time $\tau$ is set to 15 days.

### 2.3.3. Multi-Source Data Fusion Experiments

In fact, the information fusion of observations and model simulations can improve the uncertainty of single information. Therefore, combating the uncertainty of multi-source data in fusion is a key challenge of the DeepDA. The DA scheme in this experiment is the same as OSSE, which combines observation and background to generate the optimal analysis field. Different from OSSE, this experiment no longer requires setting random errors. At the same time, background fields are also provided by model simulations rather than persistent predictions. In addition, the ground truth remains the SST of ERA5, which is employed as a benchmark to evaluate the effectiveness of data fusion. Among them, The SEAS5 can better evaluate the robustness of our framework. At this stage, the weights of the GenPM have been frozen and it has been trained by ERA5.

### 2.4. Model Training and Evaluation Settings

All experiments are conducted on a workstation equipped with an NVIDIA Geforce RTX 3090 Ti-24G GPU and Tensorflow. Empirically, we set the batch size to 32 and the initial learning rate to $1\times10^{-4}$. Furthermore, the adoption of strategies such as early stopping and dynamically adjusting the learning rate can speed up the training process and prevent overfitting. It is worth noting that the GenPM weights are completely frozen during the data fusion stage of the DeepDA. The latent space vector is the only variable that can be optimized. The cost function optimization of traditional DA relies on the adjoint model to calculate the gradient [8]. To simplify the optimization process, we also employ the Adam optimizer [50] provided by Tensorflow. In the DA test phase, the range of the learning rate in Adam optimizer is set to ($1\times10^{-1}$, $1\times10^{-4}$). When the optimization rate of two adjacent steps is less than 1%, the learning rate will be reduced to half of the original value until the learning rate reaches $1\times10^{-4}$. If the learning rate drops to the minimum quickly and the cost function decreases slowly, the optimization process will be terminated after continuing for 5 steps. In the experiment, the total number of optimization iterations is less than 80. In general, obvious convergence occurs after 20 iterations and the overall process takes about 15 seconds.

In order to evaluate the performance of the DeepDA, we need to compare by different evaluation metrics. We employ root mean square error (RMSE) and anomaly correlation coefficient (ACC) as evaluation metrics. Among them, RMSE can measure the relative error of the fusion results and ACC can evaluate the correlation of spatial distribution.

$$\mathrm{RMSE} = \sqrt{\mathrm{mean}((\mathbf{X}_a - \mathbf{X}_t)^2)} \tag{14}$$

$$\mathrm{ACC} = \frac{\sum_{i=1}^{N}(\mathbf{X}_a - \overline{\mathbf{X}}_a)(\mathbf{X}_t - \overline{\mathbf{X}}_t)}{\sqrt{\sum_{i=1}^{N}(\mathbf{X}_a - \overline{\mathbf{X}}_a)^2(\mathbf{X}_t - \overline{\mathbf{X}}_t)^2}} \tag{15}$$

where $N$ is the number of spatial grid points, $\mathbf{X}_a$ and $\mathbf{X}_t$ represent the analysis field and ground truth, $\mathrm{mean}()$ is the average of samples. In practice, we only compute values for nonland grids.

# 3. Results

## 3.1. Overall Performance of DeepDA Framework

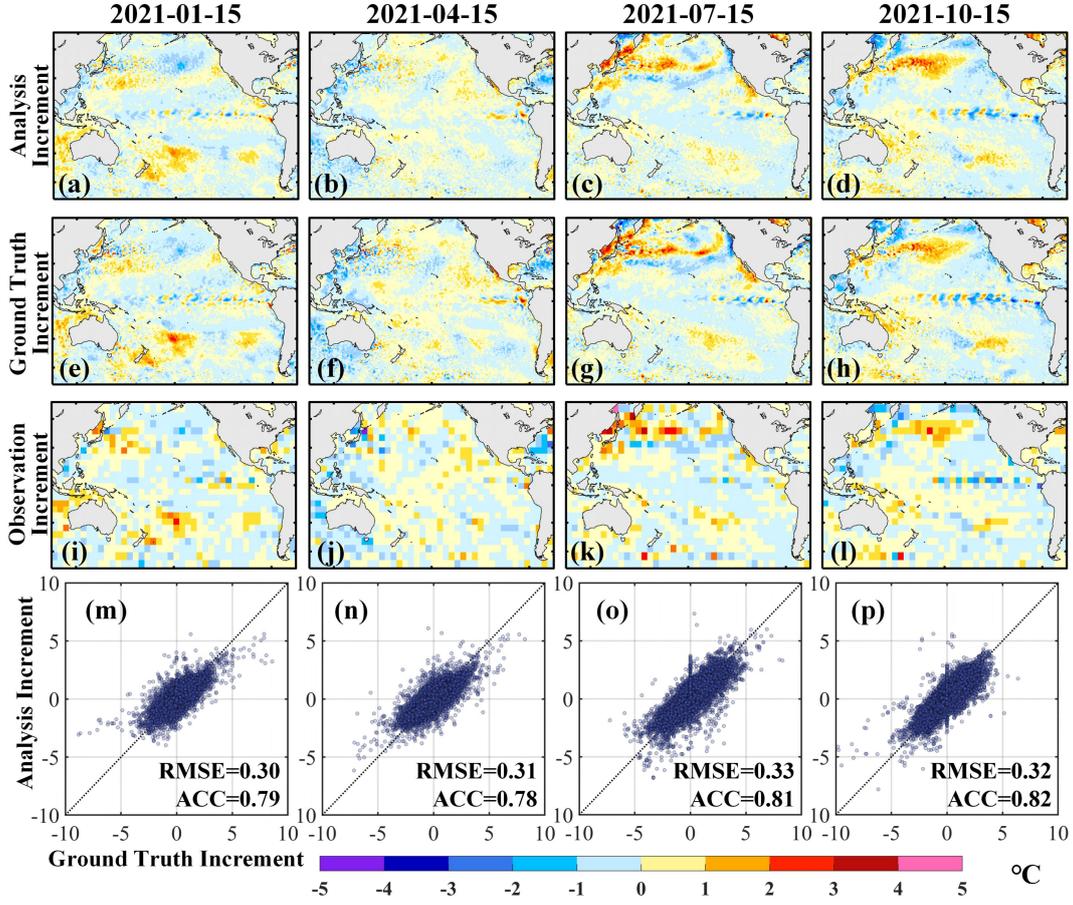

**Fig. 4.** Multi-point assimilation experiments of the DeepDA in 2021: (a)-(d) analysis increments, (e)-(h) ground truth increments, (i)-(l) observation increments and (m)-(p) regression distribution.

As shown in Fig. (4), the experiment represents the central months of different seasons in 2021. Overall, the analysis increment and ground truth increment of the DeepDA show almost the same spatial distribution. In particular, signals on larger spatial scales are effectively captured by the DeepDA. Taking July 15 as an example, the strong western boundary current from west to east in the ground truth increment (Fig. 4(g)) almost covers the North Pacific (30°N~50°N). Among them, the difference between the positive and negative increments is about 5°C. In Fig. 4(c), the pattern of positive increments is restored more accurately, which is closer to the distribution of observations. The RMSE between the ground truth increment and the analysis increment is 0.33°C, and the ACC is 0.81. In the comparison on October 15, the tropical instability wave showed a westward propagation pattern of alternating positive and negative directions (Fig. 4(h)), which is completely consistent with the analysis results. The statistical results (Fig. 4(m)-(p)) show that the RMSE of the DeepDA is around 0.3°C and the ACC is around 0.8, which is in line with expectations.

From further comparison (Fig. 4(a)-(d)), it can be found that the DeepDA weakens or even ignores the details of high-frequency small-scale signals. This is because in data fusion, the DeepDA focuses on finding the optimal state that fits the observation in the latent space, which tends to the principle that large-scale signals are preferentially fused. Other signals are often ignored because their strengths are relatively weak and the errors caused by them are random. In addition, the resolution of the observations may not be enough to fully resolve the evolution of small scales, which may also lead to certain differences between the details of the fusion and the observations. Although the refined structure cannot be completely restored, the DeepDA can provide a probabilistically optimal fusion. Its approximate distribution pattern has been accurately represented, which can enhance the data fusion potential of the DeepDA.

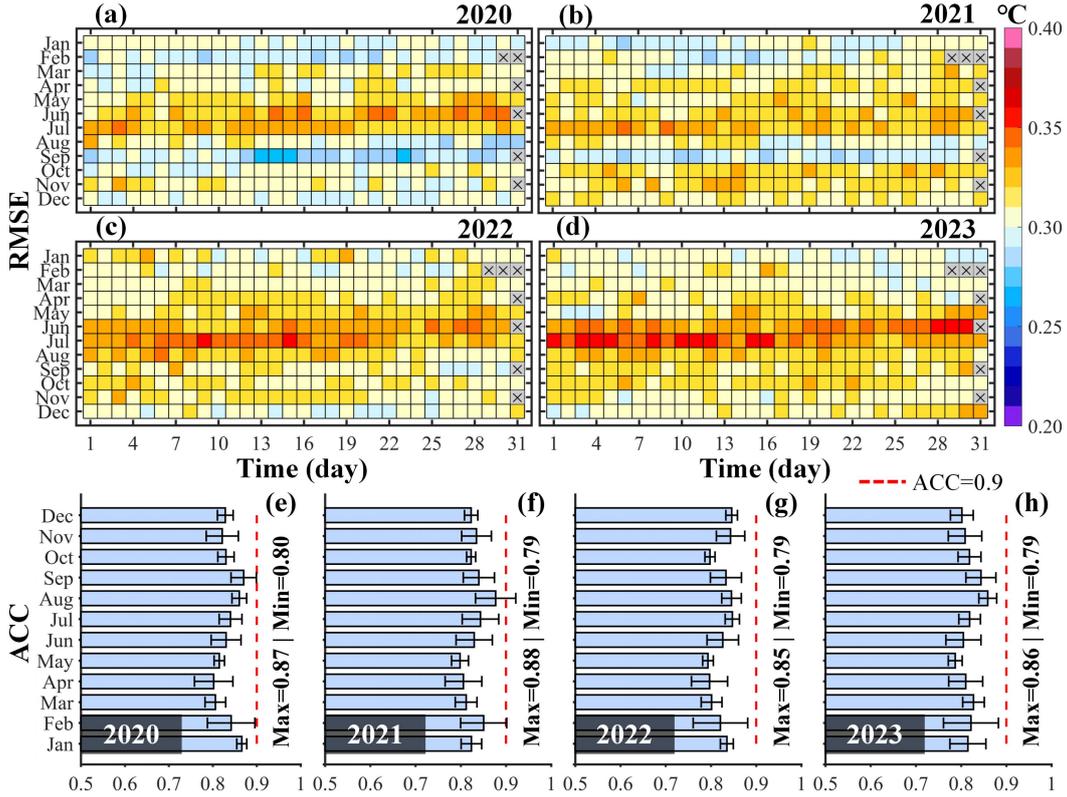

**Fig. 5.** Overall statistical performance of the DeepDA in multi-point assimilation experiments from 2020 to 2023: (a)-(d) root mean square error (RMSE) and (e)-(h) spatial anomaly correlation coefficient (ACC).

In order to further evaluate the performance of the DeepDA, a total of 1461 (4 years) DA experiments are performed and the RMSE and ACC relative to ERA5 are calculated respectively. In Fig. 5(a)-(d), the analysis RMSE of DeepDA is arranged according to the schedule. It can be seen that the overall analysis error is less than 0.4°C, mostly concentrated between 0.30°C and 0.35°C. It is worth noting that the largest range of analysis error changes is less than 10%. In the comparison of ACC, the spatial similarity between the monthly average analysis and ERA5 is stable at around 0.8. The above results all show that the DeepDA is effective and stable. More importantly, the inference time of the GenPM is almost negligible (the minimization process takes less than 1 minute), which is lightweight and efficient.

## 3.2. Generation of Unseen Nonlinear Structures

Due to low computational cost, DeepDA can effectively generate statistically significant results by increasing the number of background information samples. In the DA stage, we construct the ensemble background by adding different perturbations to the original background information. Taking 5 members as an example (Fig. S1), the SSTA on January 1, 2020 is used as the original background information, and the analysis time is April 1, 2020. In this experiment, the observation distribution has been downsampled to 2.5 degrees. As shown in Fig. S1, the analysis ACC of the five members are all above 0.9, which is significantly adjusted compared to the background information. The ensemble average result is better than the analysis of any one member, with RMSE reduced by about 30% and ACC increased to 0.95.

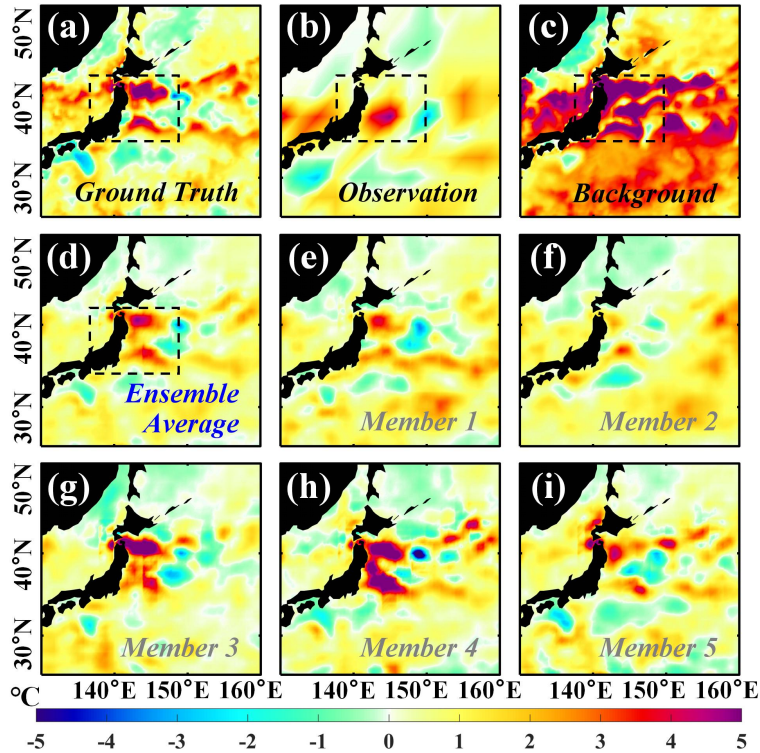

**Fig. 6.** An enlarged detail in sub-region 1: (a) ground truth, (b) observation, (c) background, (d) analysis of ensemble average, (e)-(j) analysis of members.

In fact, the distribution of observations is very sparse, so it is difficult to fully reproduce unseen nonlinear evolution processes. Fig. S1 show that the analysis results generated by DeepDA are roughly consistent with the ground truth, but there are certain differences in local details. To explore more details, we select the results of sub-region 1 to preliminarily analyze the nonlinear structure generated by DeepDA. As shown in Fig. 6(a), the nonlinear double-peak structure becomes a blurred single peak in observation (Fig. 6(b)). But DeepDA can capture the optimal spatial structure based on sparse observations (Fig. 6(d)). Similar results can be found in all members (Fig. 6(e)-(i)), which shows the excellent performance of DeepDA.

It can also be found that the diversity of background information makes the subtle structures among the analysis members similar but not completely the same. In fact, the ensemble background information can essentially expand the optimization space [51] of DeepDA, thereby reducing the uncertainty of the analysis results generated by DeepDA. Therefore, the analysis results after ensemble averaging are more statistically significant. As shown in Table S1, we compare the effects of the number of different ensemble members on DeepDA performance (sensitivity analysis). When the number of members is large, the ensemble average results are more advantageous. Therefore, different background samples in latent space DA can improve the diversity of data fusion, which can quantify the uncertainty of nonlinear systems.

### 3.3. Performance Comparison at Different Resolutions

In the previous section, we have gained a preliminary understanding of the DeepDA in generating unseen nonlinear structures. To further verify the performance, we compare the generation results guided by observation information of different resolutions. We set a fixed observation error (random seed) so that the analysis results produced by DeepDA are only affected by the resolution of the observation information. In other words, the resolution of the observation information received by DeepDA is the main controlling factor. As shown in Fig. 7 and Fig. 8, we downsample the 0.25° noisy pseudo observations to 0.5°, 1° and 2.5°, respectively.

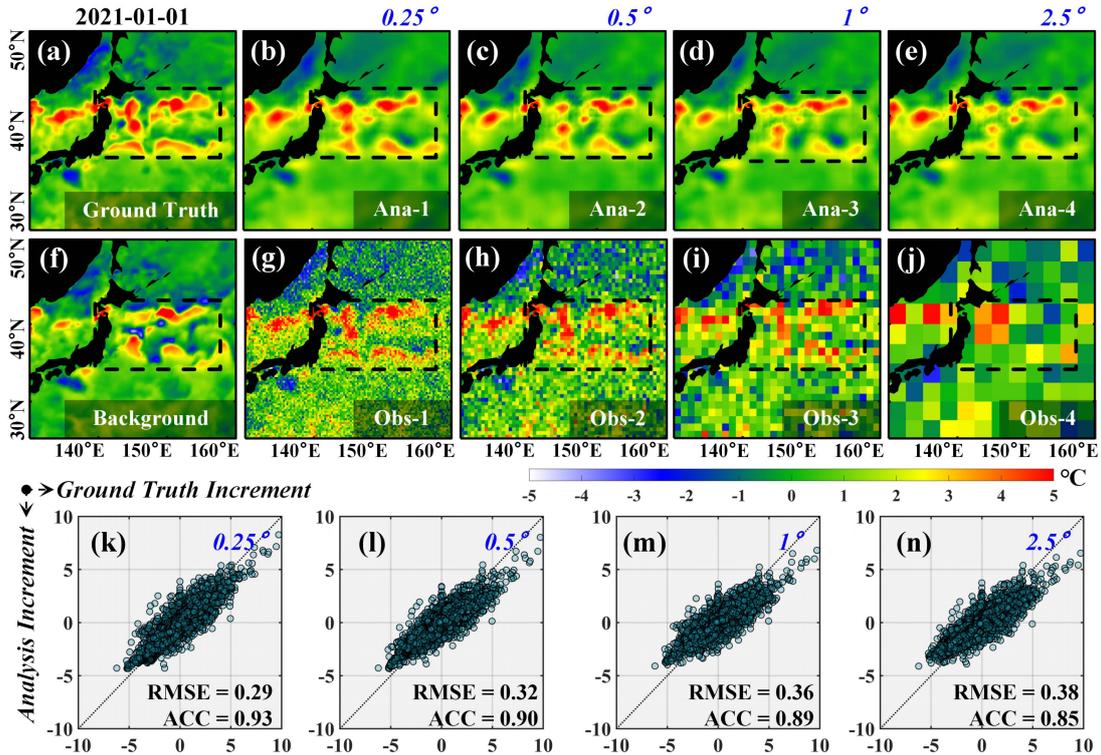

**Fig. 7.** Comparison of the (b)-(e) generation details of sub-region 1 guided by (g)-(j) observations of different resolutions (0.25°/0.5°/1°/2.5°). (a) Ground truth on January 1, 2021. (b)-(e) Analysis (Ana-1~Ana-4) results generated by DeepDA. (f) Prior background information. (g)-(j) Observation (Obs-1~Obs-4) information (noise included) at different

resolutions. (k)-(l) Regression distribution of ground truth increments (ground truth minus background) and analysis increments (analysis minus background).

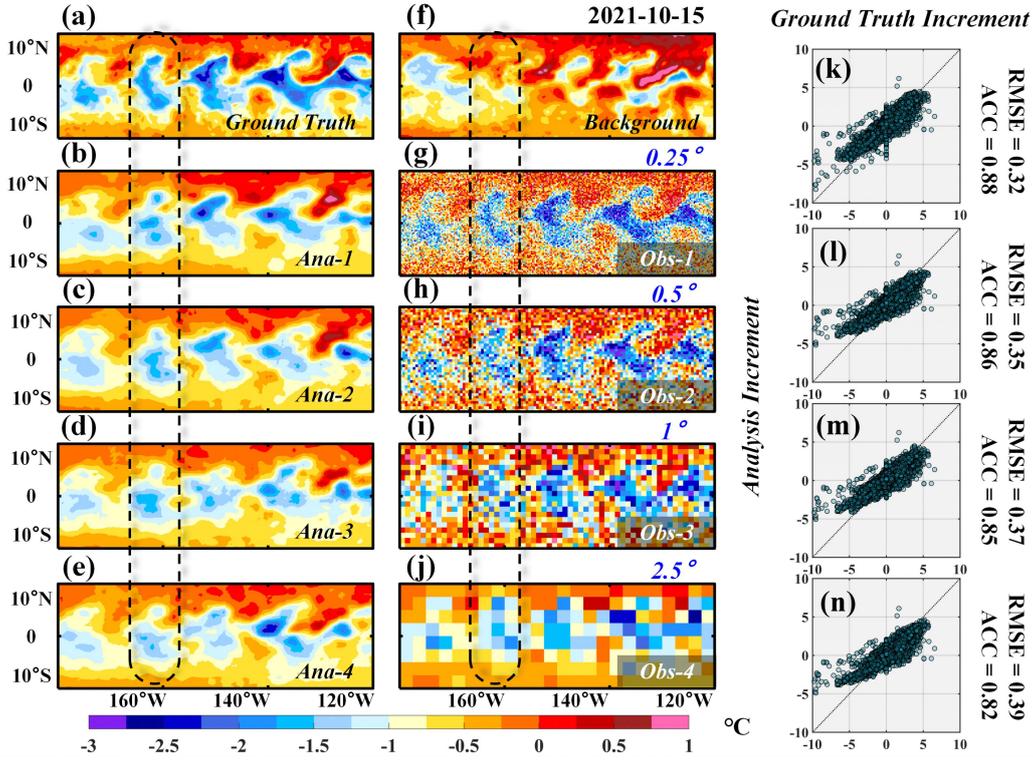

**Fig. 8.** Same as **Fig. 7**, but for sub-region 2.

Overall, the error of the analysis results generated by DeepDA increases as the resolution decreases. In sub-region 1 (Fig. 7(k)-(n)), the RMSE increases from 0.29°C to 0.38°C. In sub-region 2 (Fig. 8(k)-(n)), the RMSE increases from 0.32°C to 0.39°C. It can be found that when the available observation information (Fig. 7(j)) is only 10% of the original (Fig. 7(g)), the increase of RMSE does not exceed 40%. In other words, DeepDA can generate reasonable nonlinear evolution with only very little observation information. In detail, the complete spatial structure has been destroyed in low-resolution observations (Fig. 7(g)-(j)). Some fine structures can still be seen in the Obs-1 (0.25°), but only fuzzy or even erroneous structures can be seen in the Obs-4 (2.5°). DeepDA can reconstruct the approximate spatial structure (Fig. 7(b)-(e)) using only blurred pixel information. As shown in Fig. 8, the tropical instability wave is the core structure of sub-region 2. In the example of October 15, 2021, the wave signal propagating westward is very obvious in the ground truth. But in the sparsest observation (Obs-4, 2.5°), the wave signal is aliased with noise. It is gratifying that DeepDA can reconstruct this propagation pattern. Although the analysis results weaken the signal strength, they are still closer to the ground truth than observations and background.

By comparison, it can be found that DeepDA is robust in generating nonlinear evolution under the influence of observations at different resolutions. The generated analysis results (Ana-1~Ana-4) are quite similar in distribution. In particular, DeepDA effectively captures the nonlinear structure of different regions. In the experiments of sub-regions 1 and 2, DeepDA overcomes the artificial corruption of regional nonlinear signals. In addition, the stability of the GenPM is further verified. Sparse and noisy observations will introduce errors to the inference process of the GenPM, but the reasonable temperature structure is still captured by DeepDA. In summary, the data assimilation method in latent space can effectively generate unseen nonlinear evolution.

### 3.4. Ablation Study of Generative Proxy Models

The GenPM is the core component of the DeepDA framework, which directly affects the generation quality of nonlinear evolution. We compare the STAVAE model with 7 related models. Among them, Auto-Encoder (AE) and Variational AE (VAE) are two classic model structures and are also the root models of the STAVAE. In addition, in order to verify the effectiveness of the STAR module, we combine VAE, residual block and CBAM to build three extended models of ResVAE, VAE-CBAM and ResVAE-CBAM respectively. Details are shown in Table 1. Specifically, all hyper-parameters and layer structures are inherited directly from the STAVAE, unless otherwise specified. It is worth mentioning that we employ two different network structures: fully connected neural network (FCNN) and convolutional neural network (CNN) to construct AE (FC-AE and Conv-AE) and VAE (FC-VAE and Conv-VAE).

**Table 1.** Ablation experiments of our STAVAE model. The performance of the proxy model is evaluated based on the RMSE, ACC and model training time when reconstructing the SST.

| Model no. | Model name | Network structure | RMSE (°C) | ACC | Training time (hour) |
| --- | --- | --- | --- | --- | --- |
| 1 | FC-AE | FCNN | 2.537 | 0.263 | 1.21 |
| 2 | Conv-AE | CNN | 1.731 | 0.305 | 0.55 |
| 3 | FC-VAE | FCNN | 1.018 | 0.447 | 0.86 |
| 4 | Conv-VAE | CNN | 0.752 | 0.671 | 0.58 |
| 5 | ResVAE | CNN | 0.659 | 0.794 | 0.75 |
| 6 | VAE-CBAM | CNN | 0.602 | 0.809 | 0.71 |
| 7 | ResVAE-CBAM | CNN | 0.335 | 0.931 | 0.79 |
| 6 | **STAVAE** | CNN | 0.241 | 0.933 | 0.84 |

As shown in Table 1, the RMSE and ACC of the STAVAE (STAVAE) in the DeepDA are 0.241°C and 0.933 respectively. Compared with other proxy models, STAVAE has smaller errors and higher accuracy, which can show the potential for better generation and reconstruction. The CNN-based model is significantly better than the FCNN-based model. At the same time, the feature extraction capability of FCNN is weaker than that of CNN and there are a large number of redundant parameters. The training time of Conv-AE and Conv-VAE is reduced by about 50%, and the effect of feature learning is obvious. Considering that the input/output dimensions are close to 400,000, the structure of CNN can be faster with the support of hardware (GPU/TPU). The performance of VAE can be improved by adding residual blocks (ResVAE) between CNN layers. Residual blocks are effective for extracting multi-scale features, but it is difficult to capture spatio-temporal dependencies. VAE-CBAM introduces the attention mechanism into VAE, which captures the dependence of features, but it is difficult to fuse feature information at different scales. Therefore, the performance of ResVAE and VAE-CBAM is similar. ResVAE-CBAM integrates the above advantages and its performance has also been significantly improved. In particular, the training cost of ResVAE-CBAM is friendly to general computing hardware.

The STAVAE is a further improvement of ResVAE-CBAM, which employs the STAR module to replace CBAM. The STAVAE outperforms the above model on all three metrics, with RMSE falling below 0.3°C. The training time of STAVAE is also acceptable for our task. In summary, compared with the classic VAE and other improved models, STAVAE has good spatiotemporal representation capabilities. The STAR module can effectively improve the feature extraction capability and reduce the error of STAVAE. The excellent performance of STAVAE ensures that the DeepDA framework can capture unseen nonlinear evolutions in different application scenarios.

## 4. Discussions

### 4.1. Robustness Assessment with Real Observations

In this section, we further evaluate and discuss the robustness of the DeepDA framework. As shown in Section 3.2, rich background information can help DeepDA generate more reliable nonlinear evolution. According to the scheme benchmark in Section 2.3.3, we employ 51 members of SEAS5 as background information. SEAS5 adds perturbations to the initial conditions and drives the dynamic model to generate different prediction samples, which can approximately quantify the uncertainty of nonlinear evolution. The observational information comes from daily records provided by OISST. To match the scale of the evolution information of SEAS5, we downsample OISST to 1°.

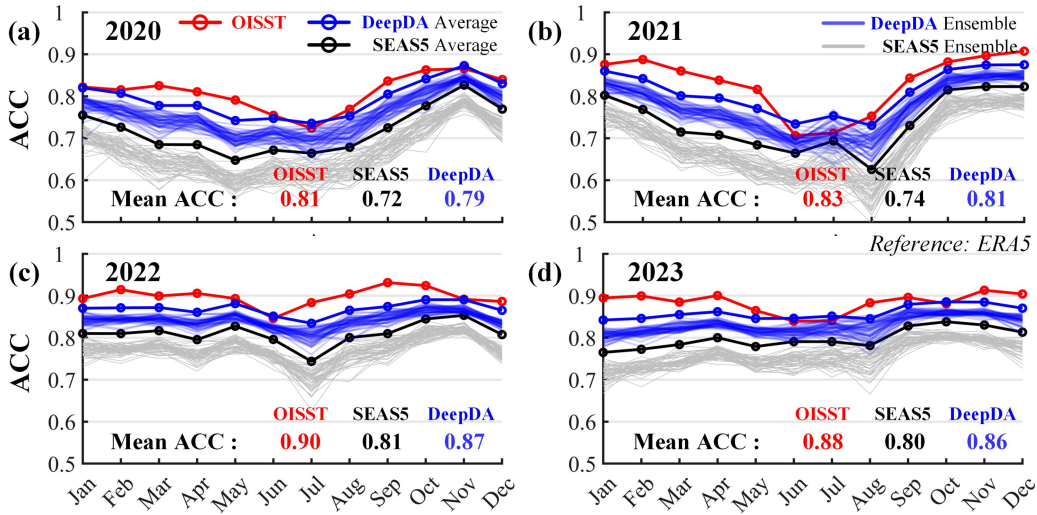

**Fig. 9.** Robustness evaluation of the DeepDA in multi-source data fusion of observations (OISST) and ensemble simulations (SEAS5). Comparison of ACC from (a)-(d) 2020 to 2023 (ground truth: ERA5).

As shown in Fig. 9, the generation results obtained by the DeepDA are significantly better than the ensemble members of SEAS5. By comparison, it can be found that the ensemble average is generally better than the simulation of a single member, which is consistent with the conclusion in Table S1. In fact, perturbations in the initial conditions will cause an increase in uncertainty, but the ensemble average can eliminate this effect (the ACC of the black line is higher than most of the gray lines and the blue line is similar). The ACC of the OISST observation and the SEAS5 ensemble average differs by about 10%, but the result of the DeepDA only differs by about 3% (Fig. 9(a)-(d)). Therefore, the DeepDA is advantageous and robust in multi-source information fusion, which means that it has greater potential in generation and data fusion. In addition, the experiment proves that background and observation information jointly affect the generation effect of nonlinear evolution, which is consistent with the experience of data assimilation. We can still see the usability of DeepDA at different time scales (daily or monthly) from the results. This shows that data fusion guided by variational assimilation constraints will be more robust and practical.

4.2. Signal Capture of ENSO Events

In order to explore the nonlinear mapping relationship established between the latent space and the physical space, we first analyze the multi-scale signal features captured by DeepDA. The GenPM can project the physical space into a low-dimensional space, which is similar to the purpose of the empirical orthogonal function (EOF) method [52]. We employ daily OISST products from 1995 to 2019 to re-drive our trained (ERA5-based) GenPM. By extracting the latent vectors in the GenPM, it can be found that the deep learning model can successfully capture the large-scale interdecadal signals in SST.

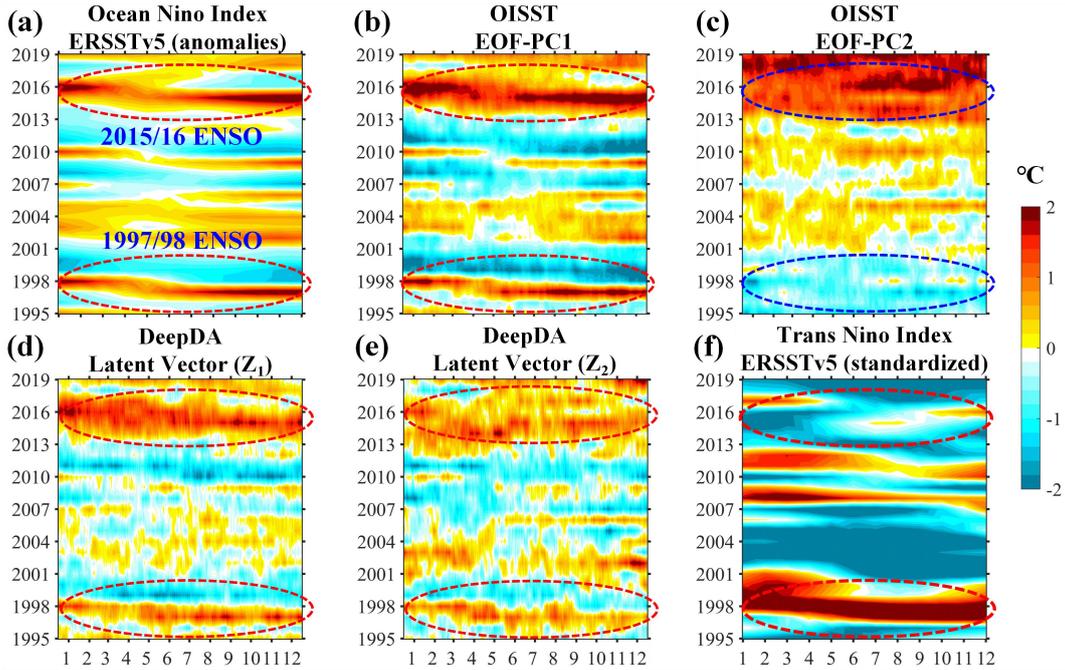

**Fig. 10.** Signal capture of ENSO events (1995-2019) by GenPM in DeepDA. (a) Ocean Niño Index from ERSSTv5. The (b) first principal component (PC1) and (c) second principal component (PC2) from OISST. The (d) first element ($Z_1$) and (e) second element ($Z_2$) of the latent vector in DeepDA. (f) Trans-Niño Index (TNI) from ERSSTv5.

As shown in Fig. 10(a) and (f), we mark two significant ENSO events [53,54] in the ocean Niño index calculated from the ERSSTv5 (NOAA Extended Reconstruction SSTs Version 5) product. We can detect the same signal in the first principal component (EOF-PC1, Fig. 10(b)) of the OISST product. It is gratifying that DeepDA can also capture similar ENSO signals (Fig. 10(d)). Different from the linear projection method represented by EOF, the deep learning model employed by DeepDA can construct a nonlinear mapping process. Therefore, nonlinear modes do not necessarily satisfy the orthogonality presented by EOF. By comparison, we can find that the signal changes controlled by the EOF modes are independent of each other (Fig. 10(b) and (c)). However, the signal changes in the GenPM are not controlled by only one mode. This conjecture can be verified from the signal distribution in Fig. 10(e). It is worth mentioning that relying on a single linear mode cannot fully describe the evolution of different scales in the ocean. The ENSO signal is captured in both the first ($Z_1$) and second ($Z_2$) dimensions of the D latent space in Fig. 10, reflecting that the same event may be jointly described by different nonlinear modes, which increases the ability to generate multi-scale evolution.

### 4.3. Explainability and Mechanism from the Pattern Perspective

In the previous section, we extracted nonlinear evolution signals that are consistent with physical information from the latent space of the GenPM. It is difficult to make deep learning models completely transparent, but we can try to obtain some explainability and mechanism of the generated results from the pattern perspective.

As shown in Fig. 11(a), the first pattern (Pattern-1) of the EOF reflects the classic ENSO signal. The ENSO signal extends over the tropical and subtropical regions. After the OISST product drives the GenPM, we extract the latent patterns in the GenPM (extraction method see Appendix B). The SST patterns corresponding to the first element ($Z_1$) and the second element ($Z_2$) of the latent vector are shown in Fig. 11(d) and (e). Consistent with our previous conjecture, both potential patterns reflect a spatial structure similar to the ENSO signal. It should be noted that the latent pattern is not globally fixed like the EOF pattern. We average the patterns of the same month, which show a consistent distribution in the spatial structure. In detail, $Z_1$ controls the positive structure (Fig. 11(d)) of the El Niño signal, which is similar to EOF Pattern-1 (Fig. 11(a)). The reverse structure (Fig. 11(e)) controlled by $Z_2$ is not symmetrical with $Z_1$, which may be related to the La Niña phenomenon. Therefore, our deep learning model can divide and conquer the two phases, which together construct the ENSO signal, but EOF (linear method) cannot fully distinguish them. The same conclusion can also be verified from Fig. 11(b) and (c), where we show the phase space constructed by the principal components (or latent vectors). We can find that $Z_1$ and $Z_2$ in 2015/16 are basically located in the first quadrant, which indicates that latent pattern 1 (positive phase) and pattern 2 (negative phase) may be mixed and inhibit each other. However, in the results of 1997/98, $Z_1$ and $Z_2$ are opposite, which shows that the two patterns reinforce each other. A previous study [55] has found that the 2015/2016 event is the strongest mixed type of El Niño ever recorded, whereas the 1997/1998 event is the strongest pure EP (Eastern Pacific) type of El Niño. In addition, the trend of SSTA in the 2015/16 event [56] undergoes a clear shift (from negative to positive).

The above results are fully consistent with our latent pattern-based analysis, which shows that our GenPM has good explainability. Compared with the EOF method, the features extracted by the deep learning model are more concentrated (Fig. 11(c)), which is essential for the generation of nonlinear evolution. In summary, the mapping of our method in latent space is closely related to the physical patterns. We have found that different latent patterns can control the generation mechanisms of different regions (Fig. S2). However, it is beyond the scope of this study to delve into this connection and provide physical validation, so this task will be a direction for future exploration.

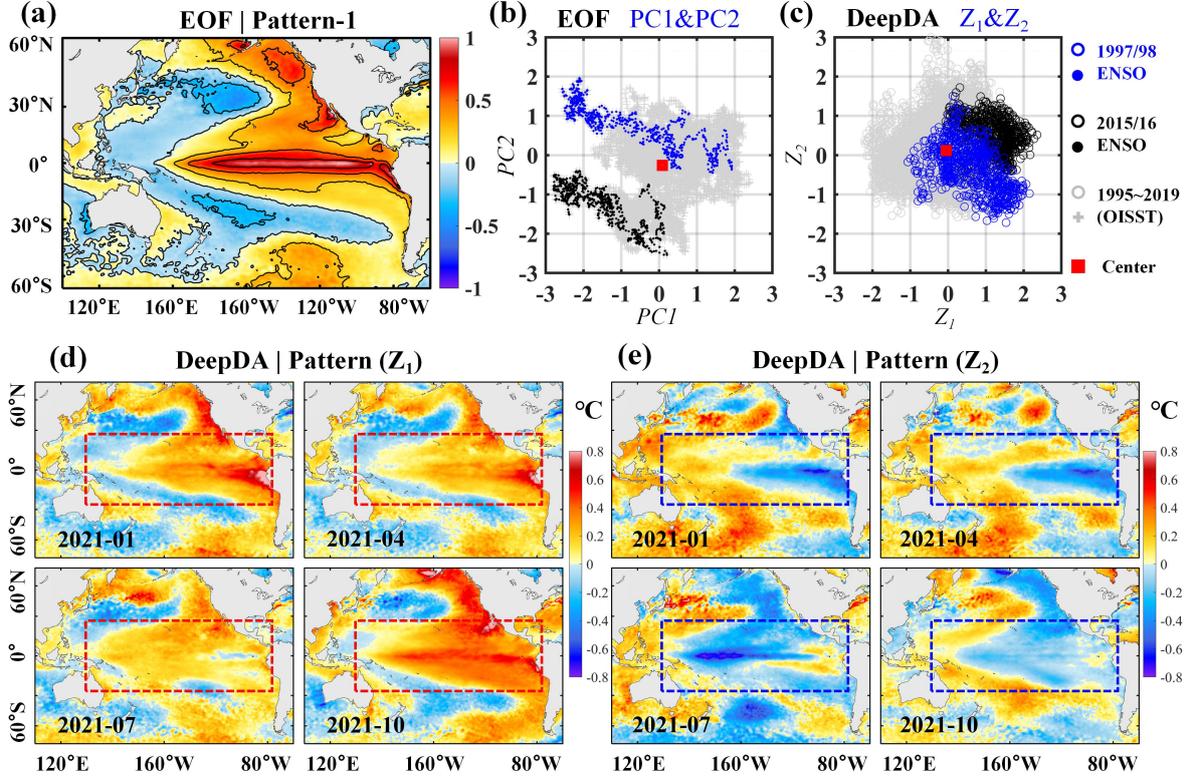

**Fig. 11.** Pattern analysis of DeepDA. (a) The first pattern of EOF (linear method). (b) The phase space represented by the first (PC1) and second (PC2) principal components of EOF. (c) The phase space represented by the first element ($Z_1$) and second element ($Z_2$) of the latent vector. (d) Nonlinear patterns corresponding to $Z_1$ and $Z_2$ in the latent space in 2021 (January, April, July, and October).

## 5. Conclusions

This study proposes a general deep learning data assimilation framework in latent space, the DeepDA. This method provides a lightweight solution for generating unseen nonlinear evolution. All experimental results show the effectiveness and stability of DeepDA in data fusion and generation. It is worth mentioning that DeepDA shows advanced data fusion skills under observations of different resolutions. Experiments with real observations have also further evaluated the robustness of DeepDA. Overall, the DeepDA is more efficient, more suitable for multi-source data fusion and can avoid large-scale computing consumption. In addition, the GenPM (STAVAE) we designed has the best performance in all aspects when compared with other deep learning baseline models. To make the study more convincing, we analyze our method from a pattern perspective. We explore the interpretability of our deep learning model from the perspective of temporal evolution and spatial distribution. Nonlinear patterns in the latent space are closely related to phenomena such as ENSO, which makes the generation results of nonlinear evolution have a relatively reasonable physical basis. In fact,

DeepDA is a new data fusion paradigm, and its performance can be further enhanced when applied to large-scale and high-quality data. It has potential in a large number of application scenarios such as Earth system prediction and satellite data fusion. It is foreseeable that the combination of DL and DA will accelerate the update of information fusion methods and products in Earth science, which will also guide our future work.

**Data availability**

We thank Google for providing the open-source deep learning framework TensorFlow (https://www.tensorflow.org) for our experiments. The ERA5 product for SST can be freely accessed (at https://cds.climate.copernicus.eu/cdsapp#!/dataset/reanalysis-era5-single-levels?tab=form). The OISST product for daily SST is freely available (at https://www.ncei.noaa.gov/products/optimum-interpolation-sst). The SEAS5 product is available for free access (at https://cds.climate.copernicus.eu/cdsapp#!/dataset/seasonal-monthly-single-levels?tab=form). The Niño SST indices can be freely accessed (at https://climatedataguide.ucar.edu/climate-data/nino-sst-indices-nino-12-3-34-4-oni-and-tni).

**Acknowledgments**

This work was supported in part by the National Natural Science Foundation under Grants 42376190, 41876014 and 42406191, in part by the National Key Research and Development Program under Grants 2023YFC3107801, 2022YFC3104800 and 2021YFC3101500.

**CRediT authorship contribution statement**

**Qingyu Zheng:** Writing-original draft, Visualization, Validation, Software, Methodology, Conceptualization. **Guijun Han:** Project administration, Funding acquisition. **Wei Li:** Methodology, Funding acquisition, Writing - review & editing. **Lige Cao:** Supervision, Methodology, Conceptualization. **Gongfu Zhou:** Formal analysis. **Haowen Wu:** Validation. **Qi Shao:** Writing - review & editing, Data curation. **Ru Wang:** Visualization, Investigation. **Xiaobo Wu:** Visualization, Investigation. **Xudong Cui:** Data curation. **Hong Li:** Writing - review & editing. **Xuan Wang:** Project administration.

---

**Appendix A. Background Error Covariance Matrix Represented in Latent Space**

The estimation of background error covariance matrix is the key in variational data assimilation. It not only represents the weight of the background, but also determines the information transfer and balance between the observation and the background.

$$\mathbf{B}_z = \left\langle (\mathbf{z}^t - \mathbf{z}^b)(\mathbf{z}^t - \mathbf{z}^b)^T \right\rangle \tag{A1}$$

The calculation method of $\mathbf{B}_z$ is shown in Eq. A1, where $\mathbf{z}^t$ and $\mathbf{z}^b$ represent the latent vectors of the ground truth and background respectively. Fig. A1 shows the absolute value of $\mathbf{B}_z$. It can be seen that $\mathbf{B}_z$ a quasi-diagonal (or diagonally dominant) matrix. The diagonal elements in $\mathbf{B}_z$ are several orders of magnitude larger than the off-diagonal elements. The estimation of $\mathbf{B}_z$ is more accurate because the number of samples used for estimation is greater than the number of elements. However, in the original estimation of **B**, the number of samples is much lower than the number of elements, which is bound to cause false correlation. In fact, the $\mathbf{B}_z$ of training set (Fig. A1(a)) and validation set (Fig. A1(b)) are not much different. But in our experiments, the DeepDA employs $\mathbf{B}_z$ inferred from the validation set, which can avoid information leakage of the test set.

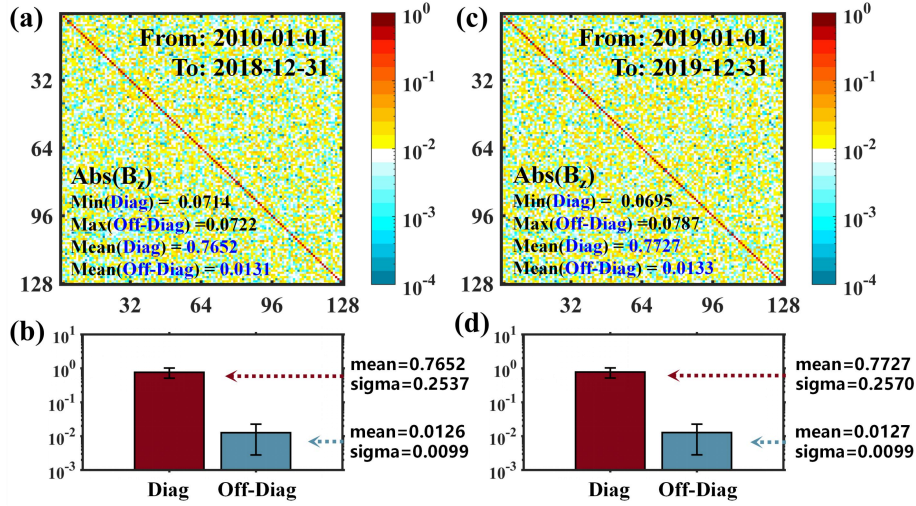

**Fig. A1.** Background error covariance matrix represented in latent space: background error covariance matrix for (a) training set and (c) validation set, and statistical tests of diagonal and off-diagonal elements for (b) training set and (d) validation set.

**Appendix B. Extracting Latent Patterns from the GenPM**

Unlike the patterns extracted by EOF, the latent patterns of GenPM determine the nonlinear mapping between the latent space and the state space. The nonlinear structure of the latent pattern is complex and contains evolutions at different scales. To visualize the nonlinear latent pattern, we want to remap it to the state space and analyze it. First, we employ the decoder of GenPM to obtain the latent vector **z** of the SST state. Then, a perturbed latent vector is generated by changing the original value of one (e.g., k-th) element of the latent vector to a different value. Finally, we employ the decoder of GenPM to map the original latent vector and the perturbed latent vector into the state space of SST. As shown in Eq. B1, the difference of the states obtained by the decoder describes the spatial structure of the latent pattern.

$$\mathbf{Pattern}(Z_k) = \text{decoder}(\mathbf{z}[k]) - \text{decoder}(\mathbf{z}[k]+1) \tag{B1}$$

In fact, the nonlinear latent patterns are different at different times. Taking Fig. 11(d) and (e) as examples, we extract the daily latent patterns for January, April, July, and October 2021 and convert them into monthly averages. It can be seen that there are similarities between different latent patterns, but they are not exactly the same, which is caused by the inherent nonlinearity of the decoder.

# Supporting Information for

# Generating Unseen Nonlinear Evolution in Sea Surface Temperature Using a Deep Learning-Based Latent Space Data Assimilation Framework

Qingyu Zheng[1], Guijun Han[1], Wei Li[1*], Lige Cao[1], Gongfu Zhou[1], Haowen Wu[1], Qi Shao[2], Ru Wang[1], Xiaobo Wu[1], Xudong Cui[1], Hong Li[1], and Xuan Wang[1]

[1]Tianjin Key Laboratory for Marine Environmental Research and Service, School of Marine Science and Technology, Tianjin University, Tianjin 300072, China.

[2]Fujian Key Laboratory on Conservation and Sustainable Utilization of Marine Biodiversity, Minjiang University, Fuzhou 350108, China.

Corresponding author:

Guijun Han (guijun_han@tju.edu.cn), Wei Li (liwei1978@tju.edu.cn)

**Contents of this file:**

**Text S1.**

The original background information is from January 1, 2020, and the analysis time is April 1, 2020. The background perturbations are selected from the SSTA 2 days before and after the same analysis time in 2019 (from March 30 to April 3, 2019). Figure S1(a) shows the significant signals in the analysis increment, which are mainly concentrated in areas where the background is significantly different from the ground truth. This shows that the DeepDA focuses on the optimal fusion between background and observation, and the proxy model increases the attention of high-impact areas.

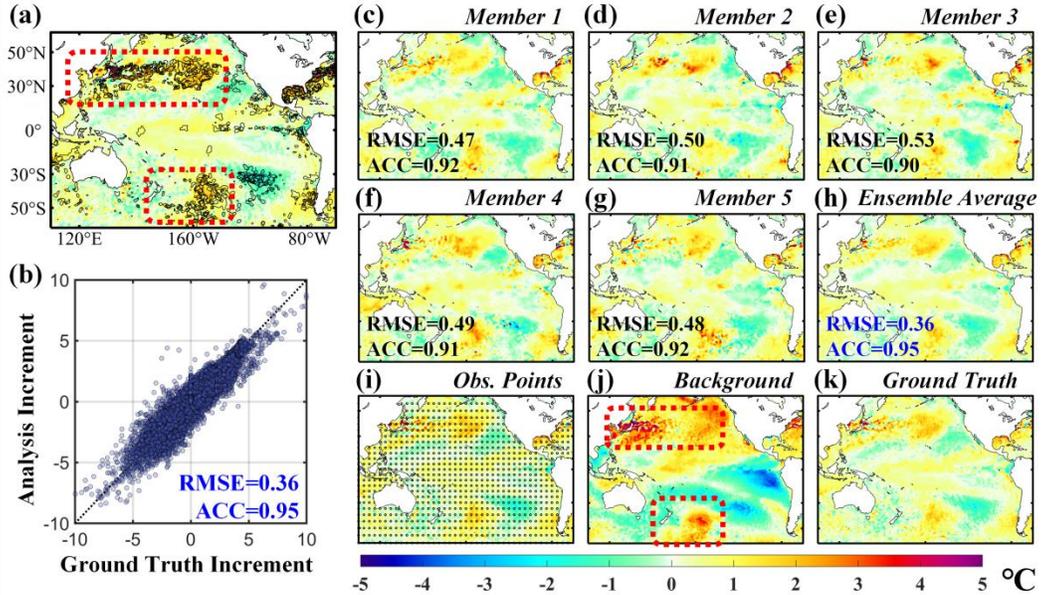

**Figure S1.** Performance evaluation of the DeepDA: (a) significant signal in analysis increments, (b) regression distribution, (c)-(g) analysis of members, (h) analysis of ensemble average, (i) layout of observation points, (j) background and (k) ground truth.

**Table S1.** Sensitivity analysis of DeepDA to different numbers of ensemble background information members.

| Number | Optimal member | | Ensemble average | |
|---|---|---|---|---|
| | RMSE (°C) | ACC (%) | RMSE (°C) | ACC (%) |
| 3 | 0.41 | 86.85 | 0.38 | 88.37 |
| 5 | 0.36 | 89.43 | 0.32 | 91.24 |
| 11 | 0.35 | 90.35 | 0.28 | 92.37 |
| 21 | 0.23 | 92.41 | 0.21 | 92.43 |

**Text S2. Spatial Patterns in Single-Point Assimilation Experiments**

We conduct three single-point assimilation experiments at different latitudes. The single observation points include a mid-latitude point in the Northern Hemisphere (45°N, 160°W), a mid-latitude point in the Southern Hemisphere (45°S, 160°W) and an equatorial point (0°, 160°W). The average annual analysis increment (analysis minus background) is shown in Figure S2. Overall, the analysis increments at the same latitude show consistent patterns. In addition, the distribution of analysis increments at different latitudes also has a typical structure. In the mid-latitudes of the Northern Hemisphere Figure S2(a)-(d), the analysis increment reaches its peak at the observation point and the strongest signal forms an east-west distribution. In the equatorial point Figure S2(e)-(h), the analysis increment shows an obvious westward propagation pattern. The overall incremental signal forms an east-west band shape as expected, which is caused by equatorial current and tropical instability waves. In the mid-latitudes of the Southern Hemisphere Figure S2(i)-(l), the positive increment pattern extends from southwest to northeast. The combined effect of westerly drifts and the south equatorial warm current can explain this phenomenon. So, this is consistent with the characteristics of ocean circulation in this area.

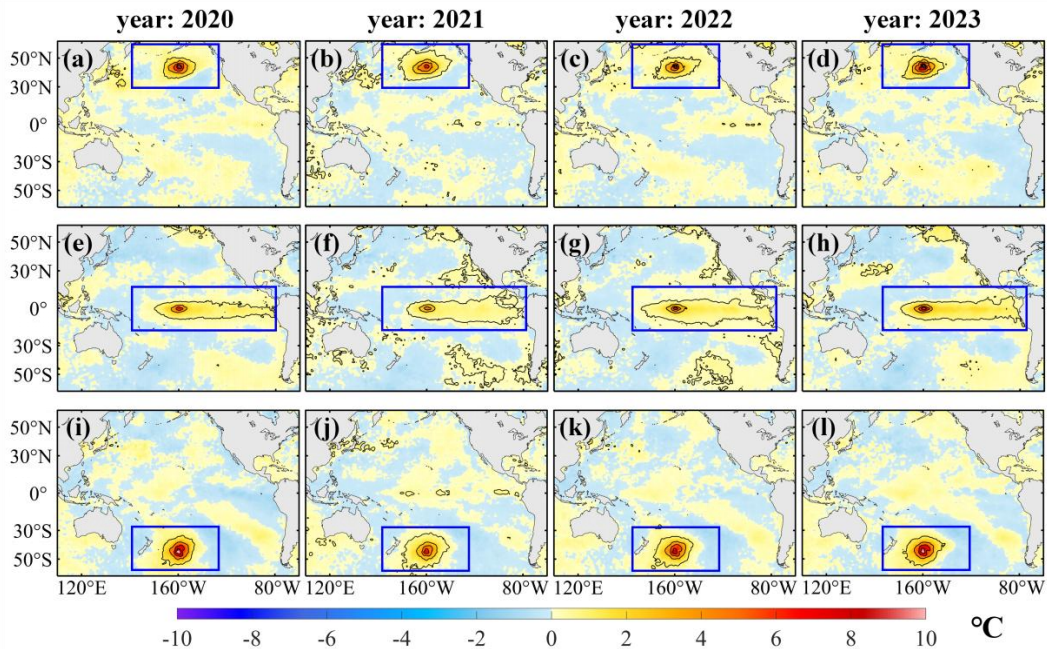

**Figure S2.** The single-point assimilation experiments from 2020 to 2023: (a)-(d) the mid-latitude point in the Northern Hemisphere, (e)-(h) the equatorial point, and (i)-(l) the mid-latitude point in the Southern Hemisphere.